\documentclass[aps,prl,twocolumn,showpacs,reprint,groupedaddress,superscriptaddress,longbibliography]{revtex4-1}

\makeatletter
\long\def\@makecaption#1#2{%
  \par
  \vskip\abovecaptionskip
  \begingroup
    \small\rmfamily
    \samepage
    \flushing
    \let\footnote\@footnotemark@gobble
    \@make@capt@title{#1}{#2}\par
  \endgroup
  \vskip\belowcaptionskip
}
\makeatother

\usepackage[english]{babel}
\usepackage{float}
\usepackage{amsfonts}
\usepackage{amssymb}
\usepackage{mathbrack}
\usepackage{siunitx}
\usepackage{mathtools}
\usepackage{bbold}
\usepackage{physics}
\usepackage{mwe}
\usepackage{amsthm,enumitem}
\usepackage{tikz}
\usepackage{ulem}
\usetikzlibrary{arrows,arrows.meta,shapes}
\usetikzlibrary{positioning}
\pdfoutput=1
\usepackage{dcolumn}% Align table columns on decimal point
\usepackage{bm}% bold math
\usepackage[utf8]{inputenc}
\usepackage{graphics,graphicx}
\usepackage{parskip}
\usepackage{amsmath,amssymb}
\usepackage{multirow}
\usepackage{fancyhdr}
\usepackage{xcolor}
\usepackage{placeins}
\usepackage[title]{appendix}
\usepackage{wasysym}
\usepackage{url}
\usepackage{braket}
\usepackage{siunitx}
\usepackage{upgreek}
\begin{document}

\title{Spin Textures and Eigenstate Evolution of Isospectrally Patterned Lattices}

\author{Peter Schmelcher}
\email{peter.schmelcher@uni-hamburg.de}
\affiliation{Zentrum f\"ur Optische Quantentechnologien, Fachbereich Physik, Universit\"at Hamburg, Luruper Chaussee 149, 22761 Hamburg, Germany}

\date{\today}

\begin{abstract}
Isospectrally patterned lattices exhibit a composite band structure with a tunable ratio of localized
versus delocalized eigenstates that is controlled by the underlying phase gradient. We show that the
lattice Hamiltonian can be interpreted as that of a single spin which, exposed to a rotating
magnetic field, hops across the lattice with a spin flip. In the low- and high-energy
part of the band, the localized states show an envelope of oscillatory character separated by quasi-nodes.
Spin peaks occur at the locations of these quasi-nodes and provide a unique spin texture to the
eigenstates, which becomes increasingly complex with increasing degree of excitation. The crossover from
localization to delocalization and vice versa leaves its fingerprints in the Fourier spectrum of the eigenstates:
the original bimodal frequency distribution widens with increasing degree of excitation, moves across
the spectral window, and finally culminates in an extremely narrow frequency peak. In the course of this
evolution, the spin texture undergoes a rearrangement transition involving different characteristic
(ir)regular patterns which we quantify by considering the total variation of the local spin fluctuations.
Our results demonstrate the variety of the spectral properties of isospectrally patterned lattices, 
which holds great prospect, in particular when considering higher lattice or cell dimensions.
\end{abstract}

\maketitle

\section{Introduction}

\noindent
Spectral degeneracies are a key ingredient and cornerstone of many quantum physical and wave phenomena. 
Their origin can typically be attributed either to the existence of sufficiently powerful symmetries
or they can be of accidental character due to the available parameter space.
Well-known examples of this are continuous rotational symmetries of atomic systems \cite{Thompson94,Hamermesh89}
and conical intersections of adiabatic molecular potential energy surfaces \cite{Koeppel84,Baer06}.
The degree of degeneracy can vary largely such that the energetically degenerate subspace
can range from low-dimensional to, in principle, infinite-dimensional, and with the possibility
of continuous parameter-dependent degeneracies of a certain codimension \cite{Koeppel84,Baer06}.
Macroscopic degeneracies and quantum states are at the heart of major quantum phenomena, such as the famous
Cooper pairing of s-wave superconductivity \cite{Tinkham04}, Bose-Einstein condensation of ultracold atoms \cite{Pethick11},
flat band occurrence in extended lattices \cite{Leykam18} or bulk systems \cite{Aoki25}.
All of these examples share the common feature that degenerate eigenstate (sub-)spaces are highly 
sensitive to perturbations such as disorder, interactions, and external field effects, thereby providing
intriguing quantum phenomena at their individual scales. Indeed, in many cases the perturbation is
of symmetry-breaking character and mixes the degenerate eigenstates, which can allow for the preparation
of specific new eigenstates with desired properties or even functionalities. This way,
strongly correlated symmetry-broken quantum many-body phases can emerge, such as fractional Chern insulators \cite{Okuma23}.

\noindent
Symmetry breaking can happen in several consecutive steps. One pathway is to break a global symmetry, such 
as a reflection/inversion symmetry, but to retain it locally, in the sense that it now holds only on a finite
region of space for a system that applies globally. This approach leads to locally symmetric setups whose theory
has been developed and applied in recent years \cite{Kalozoumis14a,Kalozoumis13a,Kalozoumis13b,Morfonios17,Morfonios20}
and has been experimentally probed in acoustic and electromagnetic wave systems \cite{Kalozoumis15,Schmitt20}.
A key observation omnipresent in locally symmetric setups is the fact that eigenstates have a strong tendency
to localize on spatial domains where the underlying potential possesses a local symmetry, such as reflection
symmetric potential domains \cite{Roentgen19}. The origin of this behavior has been traced back to the isospectrality
of the isolated symmetry-related parts of the locally symmetric region: they exhibit an exact degeneracy
and their spatial interface serves as a nucleus for the localization upon switching on the couplings \cite{Schmelcher24}.

\noindent
The above-described mechanism has been elevated to a guiding construction principle for the so-called isospectrally
patterned lattices (IPL). IPLs are composed of coupled cells corresponding to isospectral blocks located on the
diagonal of the lattice Hamiltonian. These blocks vary across the lattice but are all obtained from a single 
diagonal 'seed' block Hamiltonian by orthogonally, or generally unitarily, transforming the seed Hamiltonian.
The latter is achieved by varying the rotation angles of the transformation across the lattice. IPLs are therefore
designed by spectral degeneracy engineering: they are, in general, non-periodic inhomogeneous lattice setups.
They fall into the rich gap between (quasi-)periodic \cite{Macia09,Macia21} and random lattices \cite{Lee85}.
A first exploration \cite{Schmelcher25} has demonstrated the intriguing properties of IPL: the bands are composed
of a sequence of single center localized eigenstates energetically
 neighboring to the band edges and a sequence of delocalized states 
neighboring to the band center, which span across the complete finite lattice. The key parameter is the phase
gradient of the lattice, which allows to tune the fraction of localized versus delocalized eigenstates and determines
the localization length \cite{Diakonos26}. Recently, several alternative IPL setups have been explored where the
range of angles covered by the lattice has been varied, thereby changing the decomposition of the
spectrum and obtaining a multi-center localization as well as a symmetry breaking of the localization
behavior \cite{Schmelcher26}.

\noindent
While the IPL in the above settings shows a rich phenomenology, its concrete physical realization and interpretation
as well as experimental implementation has not been settled. In the present work, we address this gap and show that
the IPL Hamiltonian can be interpreted in terms of a single mobile spin in a rotating magnetic field 
allowed to move across the lattice by a spin-dependent hopping. Based on this interpretation, we
analyze the local spin textures of the eigenstates and find that, in the low- and high-energy sector, on top
of a smoothly varying background, strongly localized spin peaks occur at the positions of the quasi-nodes 
of the eigenstates. We identify the patterns of this quasi-nodal structures. Furthermore, it is shown that the Fourier
spectral decomposition of the eigenstates changes from two separated distributions occurring at low and high
frequencies to a single spectral peak and vice versa while moving across the entire frequency spectrum. 
Strong spin-spatial correlations and large total spin fluctuations are found with increasing degree of
excitation. 

\noindent
Our work is structured as follows. In section \ref{setup} we define our setup and the underlying Hamiltonian,
its symmetries, and in particular its interpretation in terms of a concrete physical system.
Section \ref{eigenstate} contains an
eigenstate analysis, with subsection \ref{quasinodes} focusing on the quasi-nodal structure of the
eigenstates and subsection \ref{fourier} providing insights on the structural rearrangement of the
eigenstates with increasing degree of excitation via a Fourier transformation.
In section \ref{spin} we analyze the local spin structure
of the eigenstates and put it in relation to the quasi-nodal structure. Section \ref{coupling}
briefly addresses the phenomenology for stronger couplings. Finally, we present in section
\ref{concl} our conclusions and outlook.

\section{Setup, Hamiltonian and Spin Interpretation}
\label{setup}

\noindent
Isospectrally patterned lattices \cite{Schmelcher25,Schmelcher26}
are composed of coupled isospectral cells, i.e., of individual cells 
which possess the same set of energy eigenvalues. The cells are given by $K \times K$ matrices ${\mathbf{A}}_{m}$, 
where $m$ is the cell-index in the lattice. A main feature of IPL is the fact that these degeneracies
among different cells can be systematically designed by creating the cells from the same diagonal
structure ${\mathbf{D}}$ while applying different orthogonal (or in general unitary) transformations 
${\mathbf{O}}_{{\bm{\phi_m}}}$.
The latter are parametrized by (a set of) corresponding angles ${\bf{\phi}}_m=\{\phi^1_m,...,\phi^{N_p}_m\}$
with $N_p=\frac{K(K-1)}{2}$ that vary in a patterned manner when 'moving' across the lattice. This leads to
the constituting relation

\begin{equation}
{\mathbf{A}}_{m} = {\mathbf{O}}_{\phi_m}^{-1} {\mathbf{D}} {\mathbf{O}}_{\phi_m}, \hspace*{0.5cm} m \in \{1,...,N\},
\label{eq1}
\end{equation}

\noindent
where $N$ is the number of cells of the lattice. There is a large variety and flexibility in
designing the angle sequences, which adds to the richness of IPL. We are focusing here on finite
inhomogeneous lattices covering a certain phase interval described by the Hamiltonian

\begin{eqnarray}
{\cal{H}} &=& \sum_{m=1}^{N} \left(\ket{m} \bra{m} \otimes \mathbf{A}_{m} \right) \nonumber \\ 
& + & \sum_{m=1}^{N-1} \left(\ket{m+1} \bra{m} \otimes \mathbf{C}_{m} + h.c. \right), 
\label{eq2}
\end{eqnarray}

\noindent
where the coupling between the cells is provided by 
${\mathbf{C}}={\mathbf{C}}_m = \frac{\epsilon}{2} \left(\sigma_x + i \sigma_y \right)$
with the coupling parameter $\epsilon$, and $\sigma_i, i=x,y,z$ are the Pauli matrices.
$N_s$ will, in the following, denote the
total number of lattice sites within and across cells. We will focus throughout this work
on the real case $K=2$, resulting in a single phase parameter $\phi$, and on finite lattices covering a finite
interval of the phase $\phi$ and possessing a constant phase difference between neighboring cells,
i.e., an equidistant grid of angles centered around the value $\frac{\pi}{4}$.
The complete angular (or phase) range covered by the lattice reads then
$[\frac{\pi}{4}-\frac{L}{2},\frac{\pi}{4}+\frac{L}{2}]$ with $L = \frac{\pi}{4}$,
and we have $\phi_m = \frac{\pi}{4} - \frac{L}{2} + \frac{m-1}{N-1} L, m \in \{1,...,N\}$.
Our lattice possesses, by construction, an inversion (unitary) symmetry around its center $\phi = \frac{\pi}{4}$, which
reads ${\cal{I}}= \sum \limits_{m} |m \rangle \langle N-m+1| \bigotimes {\sigma_x}$ with ${\cal{I}}^2 = {\mathbf{I}}$.
We employ open boundary conditions for our lattice.

\noindent
Specializing to the above-mentioned case, we have

\begin{equation}
\begin{aligned}
\mathbf{A}_m = &
\begin{bmatrix}
\cos \phi_m &   \sin \phi_m \\
- \sin \phi_m & \cos \phi_m \\
\end{bmatrix}
\begin{bmatrix}
d_1 &  0 \\
0 & d_2  \\
\end{bmatrix}
\begin{bmatrix}
\cos \phi_m &   -\sin \phi_m \\
\sin \phi_m & \cos \phi_m \\
\end{bmatrix}  \\
= &
\begin{bmatrix}
d_1 \cos^2 \phi_m + d_2 \sin^2 \phi_m &   (d_2 - d_1) \sin \phi_m \cos \phi_m\\
(d_2 - d_1) \sin \phi_m \cos \phi_m  & d_1 \sin^2 \phi_m + d_2 \cos^2 \phi_m \\
\end{bmatrix},
\end{aligned}
\label{eq3}
\end{equation}

\noindent
and our Hamiltonian 
(\ref{eq2})
can then be reformulated to

\begin{figure*}[t]
\hspace*{-2cm} % Adjust this value to shift the plot left or right
\includegraphics[width=19cm,height=6.5cm]{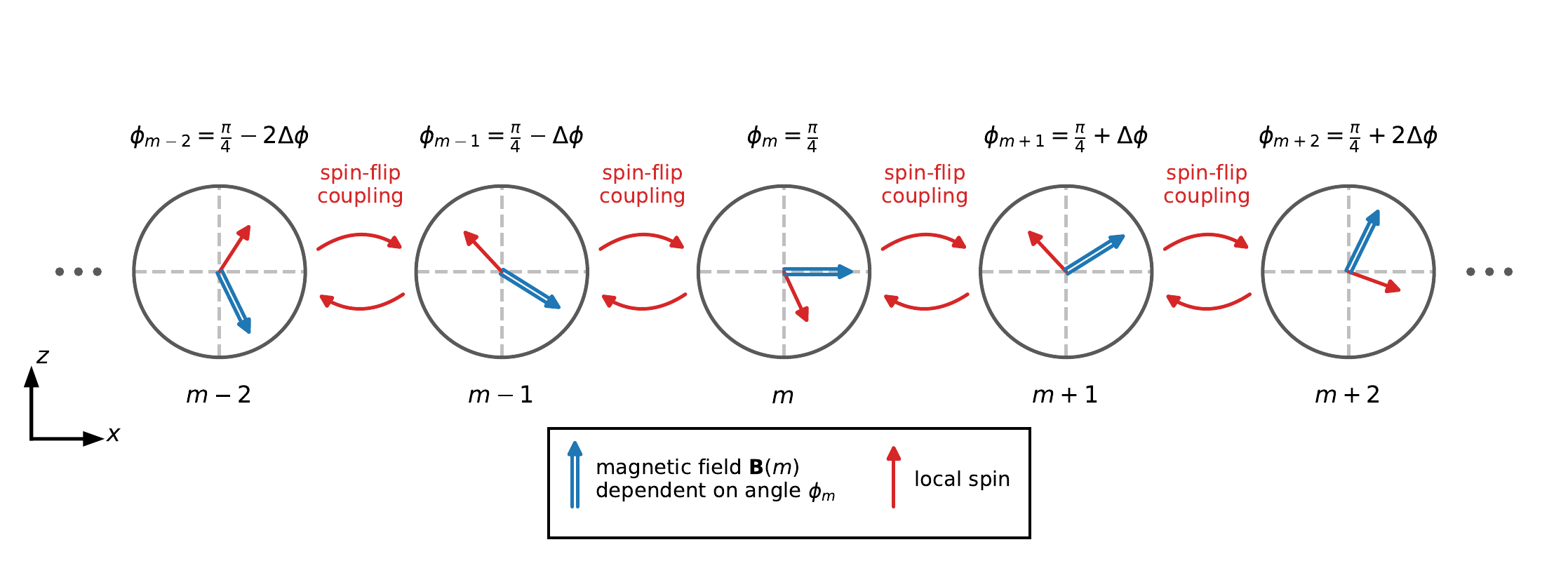}
\caption{Graphical sketch of the IPL Hamiltonian in eq.(\ref{eq5}). Indicated are the local magnetic field
${\bf{B}} (m)$ (blue vector), the local spin (red vector) as well as the spin flip
coupling between different cells. The cells are labeled by $m-2,m-1,m,m+1,m+2$.}
\label{Fig:0}
\end{figure*}

\begin{eqnarray}
{\cal{H}} &=& \sum_{m=1}^{N} \ket{m} \bra{m} \otimes \left[\left(\frac{d_1+d_2}{2}\right) \mathbb{1} \right. \nonumber \\ 
&& \left. + \frac{1}{2} \left(d_1 - d_2 \right) \left( \cos 2 \phi_m \sigma_z - \sin 2 \phi_m \sigma_x \right) \right]
\nonumber \\ 
&& + \epsilon \sum_{m=1}^{N-1} \left( \ket{m+1} \bra{m} \otimes \left(\frac{\sigma_x + i \sigma_y}{2} \right) + h.c. \right).
\label{eq4}
\end{eqnarray}

\noindent
Note that the Hamiltonian ${\cal{H}}$ has the trace $Tr({\cal{H}}) = N \cdot (d_1+d_2)$ and can be
rendered into a traceless form simply by subtracting as follows ${\cal{H}} - \frac{1}{2} (d_1+d_2) \cdot {\mathbf{I}}$,
which shifts all eigenvalues by the global amount $\frac{1}{2} (d_1+d_2)$.

\noindent
Let us now provide a physical interpretation of the IPL Hamiltonian. We regard
the two components within each cell as the two states of an internal spin-$\frac{1}{2}$
degree of freedom, and the cell index $m$ as the position on a one-dimensional
spatial lattice. In this picture, and omitting the constant shift
$\frac{1}{2}(d_1+d_2)$, the Hamiltonian (\ref{eq4}) reads

\begin{eqnarray}
{\cal{H}} &=& \sum_{m=1}^{N} \ket{m} \bra{m} \otimes \left( \alpha B_z(m) \cdot \sigma_z + \alpha B_x(m) \sigma_x \right)
\nonumber \\ 
&&+ \epsilon \sum_{m=1}^{N-1} \left( \ket{m+1} \bra{m} \otimes \sigma_+ + h.c. \right),
\label{eq5}
\end{eqnarray}

\noindent
with

\begin{eqnarray}
\alpha B_z (m) & = & \frac{1}{2} \left(d_1 - d_2 \right) \cos 2 \phi_m, \\ \nonumber 
\alpha B_x (m) &=& -\frac{1}{2} \left( d_1 - d_2 \right) \sin 2 \phi_m.
\label{eq6}
\end{eqnarray}

where $\sigma_\pm = \frac{1}{2}(\sigma_x \pm i\sigma_y)$ are the spin raising and
lowering operators, i.e., ${\mathbf{C}} = \epsilon\,\sigma_+$.

\noindent
The physical picture is therefore as follows. The IPL describes a single
spin-$\frac{1}{2}$ particle on a one-dimensional lattice, subject to two processes.
First, at each site $m$ the spin couples to a local magnetic field ${\bf{B}}(m)$
of constant magnitude $|{\bf{B}}| = |d_1-d_2|/(2\alpha)$ lying in the $(x,z)$-plane,
whose direction is determined by the angle $2\phi_m$. Across the lattice the
field direction therefore rotates by the total angle $2L = \frac{\pi}{2}$, and at
the lattice center ($\phi = \frac{\pi}{4}$) it points along the $x$-axis.
The constant $\alpha$ contains all physical constants, such as the magnetic
moment and the $g$-factor. Second, the particle hops between neighboring sites,
and every hop is accompanied by a spin flip: a hop to the right,
$m \rightarrow m+1$, raises the spin ($\sigma_+$), whereas a hop to the left,
$m+1 \rightarrow m$, lowers it ($\sigma_-$). Spin-conserving hopping is absent.

\noindent
This spin-flip hopping has an important consequence. Taken alone, it only couples
the state $|m,\downarrow\rangle$ to $|m+1,\uparrow\rangle$, i.e., it decomposes the
lattice into isolated pairs of states and does not allow for transport. A
connection to the next pair is provided by the transverse component $B_x(m)$ of
the local field, which couples $|m,\uparrow\rangle$ and $|m,\downarrow\rangle$
within the same cell. The IPL thus forms a chain in which spin-flip hopping between
neighboring cells ($\epsilon$) alternates with intra-cell spin flips ($\alpha B_x(m)$),
and the spatial variation of the field direction directly enters the
propagation of the particle along the lattice.

\noindent
The magnetic field strength components are proportional to the difference of the diagonal elements $(d_1-d_2)$.
Fig. \ref{Fig:0} provides a graphical illustration of the IPL Hamiltonian, indicating the
local magnetic field ${\bf{B}}(m)$, the spin varying from cell to cell ($m$) as well as the spin flip process 
between different cells. Based on this physical picture, an experimental 
implementation of our IPL will be discussed in the conclusions. Before we perform a spin analysis of our IPL,
see section \ref{spin}, which is motivated by the above interpretation,
let us investigate the evolution of the properties of the eigenstates with increasing degree of excitation.

\begin{figure*}[t]
\includegraphics[width=19cm,height=11cm]{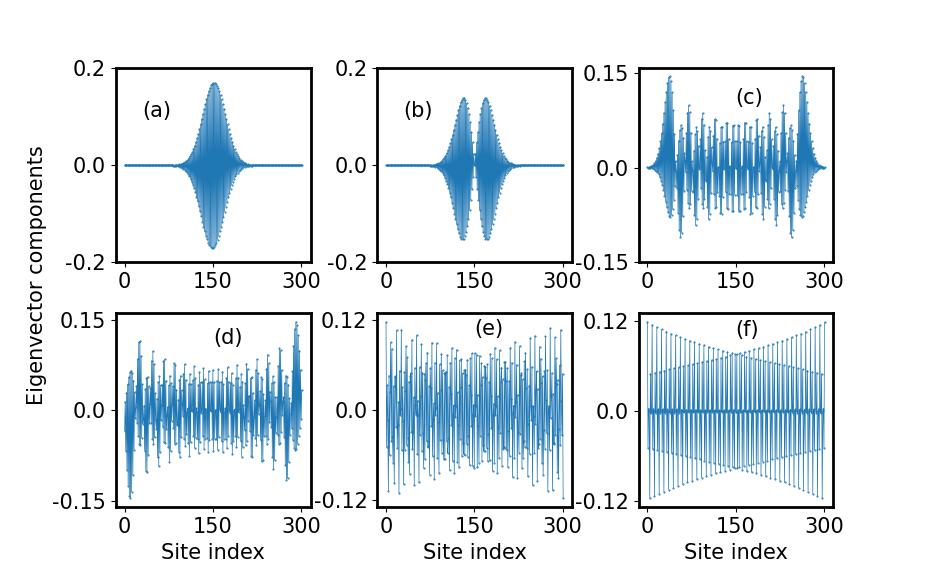}
\caption{Individual eigenstates of the IPL for $d_1=1,d_2=2, \epsilon=0.01,
N_s=302$ with open boundary conditions, placed symmetrically around $\frac{\pi}{4}$.
Shown are the values of the sequentially labeled (site index) eigenvector components accounting for the 
total eigenvector of the IPL Hamiltonian.
$\epsilon$ is the off-diagonal coupling, and $N_s$ is the dimension of the Hamiltonian. The
subfigures (a-f) correspond to the $0,1,19,30,70,75$-th eigenstate in the first band, respectively. (a-c) are low
energy localized states, and (d-f) belong to the delocalized states. The $75$-th eigenstate is at the center
of the first band.}
%}}
\label{Fig:1}
\end{figure*}

\section{Eigenstate Analysis}
\label{eigenstate}

\noindent
IPL show intriguing spectral properties \cite{Schmelcher25,Schmelcher26}.  
It has been demonstrated that each energy band consists of three subdomains with a qualitatively
different behavior of the eigenvalue spectra. Most importantly, these subdomains correspond to 
a different localization behavior of the corresponding eigenstates. For low energy eigenstates
close to the lower band edge of the first band, we observe a series of localized states which
are all centered around the center of the lattice and spread with increasing degree of excitation.
It has been shown that the origin of this localization behavior and its localization
length is due to the discrete phase gradient
in conjunction with the coupling between the cells of the lattice \cite{Schmelcher25,Diakonos26}.

\noindent
Once the widths of these eigenstates reaches the boundaries of the finite lattice, we obtain a
series of delocalized states which occupy energetically a subdomain around the center of the
band. For higher energies and neighboring to the upper band edge, again a series of localized states emerge,
and this behavior repeats for the upper band.
This is visualized in Fig.\ref{Fig:1}, where some selected localized and delocalized eigenstates
from the lower half of the lower band are shown.
Note that the significant energy difference between the low-energy and high-energy localized states
can be ascribed to the different short-range behavior of the eigenstates,
i.e. to the characteristic oscillatory variation of the amplitudes of the eigenstates from site to site.

\noindent
A note is in order concerning our use of the term (de-)localized eigenstates. Since we address in this
work exclusively finite inhomogeneous lattices, we apply the term localized eigenstates to those states which possess
a substantial amplitude solely inside the lattice and do not reach the boundaries.
Delocalized eigenstates extend over the complete lattice. Localized and delocalized states are not
intermingled but occur in specific energy regimes and are separated 
by a finite system localization delocalization crossover.

\noindent
It can be seen from Fig.\ref{Fig:1} that the localized eigenstates possess an envelope behavior, which we
shall analyze in terms of its quasi-nodal structure in the following.

\subsection{Quasi-Nodal Structure of the Eigenstates}
\label{quasinodes}

\begin{figure}[ptb!]
%\begin{minipage}[t]{0.9\textwidth}
\hspace*{-0.7cm} % Adjust this value to shift the plot left or right
\includegraphics[width=9cm,height=7cm]{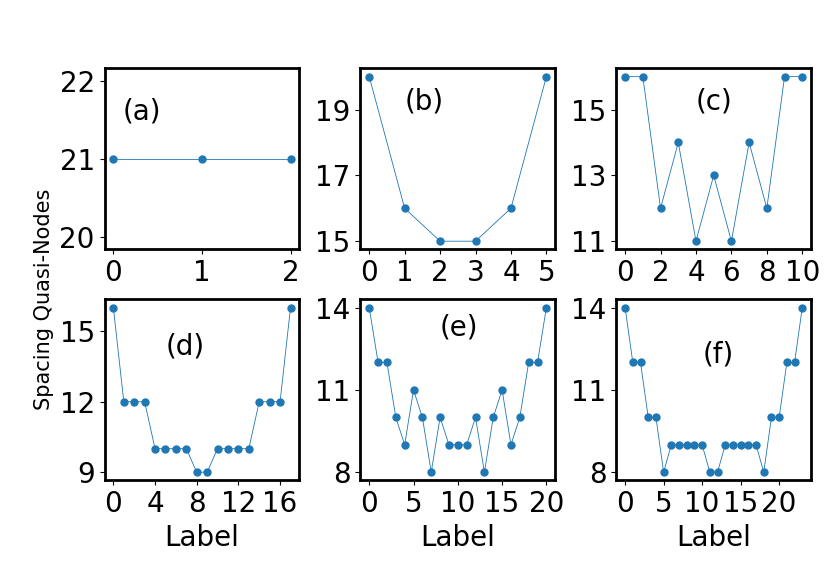}
%\end{minipage}
\caption{The lattice distances between neighboring quasi-nodes versus their consecutive labeling 
for individual selected eigenstates. (a-f) correspond to the $4,7,12,19,22,25$-th eigenstate, respectively, of
the first band. The underlying lattice is for $d_1=1,d_2=2,\epsilon=0.01,
N_s=302$ with open boundary conditions, placed symmetrically around $\frac{\pi}{4}$.
$\epsilon$ is the off-diagonal coupling, and $N_s$ is the dimension of the Hamiltonian.}
\label{Fig:2}
\end{figure}

\noindent
The envelope behavior of the eigenstates, as exemplarily demonstrated in Fig.\ref{Fig:1}, is particularly
evident for the localized states but extends into the domain of delocalized states where it eventually
loses its meaning (see subfigures \ref{Fig:1}(e,f)). Note that, for the case $N_s=302$ shown in Fig.\ref{Fig:1},
there are within the first band $25$ localized eigenstates around each band edge and $101$
delocalized states that extend over the complete finite IPL. This has to be seen as an approximate
statement since there is a crossover regime from localization to delocalization.
The individual eigenstates displayed range
from the ground state to the eigenstate in the center of the first band.
The oscillatory envelope exhibits quasi-nodes where
the eigenstates possess close to zero values. In the low energy sector, the number of quasi-nodes increases
with increasing degree of excitation, whereas the opposite holds for the high energy sector of localized
states close to the upper band edge (be reminded that their energy difference can be attributed to the short range
properties of the eigenstates). Fig.\ref{Fig:2} shows the spacing, in terms of lattice sites, between
neighboring quasi-nodes for a selection of low-energy localized eigenstates. First of all, we observe a 
reflection symmetry around the center, which goes back to the inversion symmetry of the IPL. Second, one notes
that the spacing increases with increasing node number from the center, with some oscillations on top.
Third, we encounter plateau-like features of equal spacing of the quasi-nodes.

\noindent
To explore the regular vs. irregular character of the localized versus delocalized eigenstates, we analyze
in the following subsection the spectral content of the eigenstates in terms of their Fourier transform.

\subsection{Fourier Analysis and the Evolution of the Eigenstates}
\label{fourier}

\noindent
Let us now inspect the Fourier transformation of the eigenstates, more precisely the magnitude of the
Fourier coefficients, as we 'move' across the (first) band, passing from localized to delocalized and
back to localized eigenstates. Note that, for the Fourier transformation of our IPL, there
is no scale of the lattice naturally available, and we therefore choose the sample frequency or
sampling rate $f_s = 1$, which is twice the largest frequency covered by the Fourier transform. 
We remark that the notion of frequency is here based on the lattice site arrangement.
Together with the given size of our lattice $N_s=1202$, chosen here to allow for a thorough Fourier analysis,
 this defines all resulting parameters, such
as the frequency binning $\Delta f = f_s/N_s$, which corresponds to the smallest frequency. 

\begin{figure}[ptb!]
%\begin{minipage}[t]{1.0\textwidth}
\hspace*{-0.7cm} % Adjust this value to shift the plot left or right
\includegraphics[width=9cm,height=10cm]{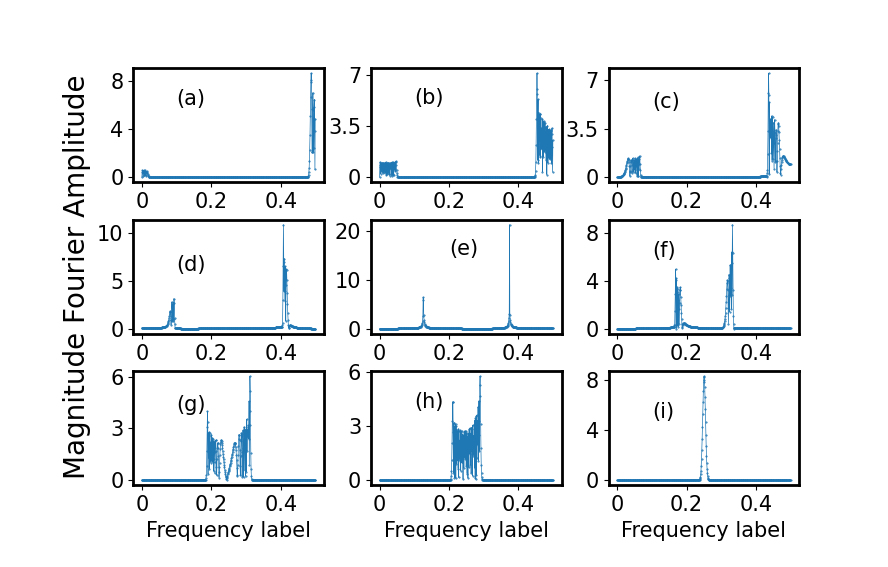}
%\end{minipage}
\caption{Absolute values of the amplitudes in frequency space for the 
individual $6,70,130,213,302,416,480,550,600$-th eigenstate corresponding to subfigures (a-i)
for an equidistant $\phi$ lattice for $d_1=1,d_2=2,\epsilon=0.01,
N_s=1202$ with open boundary conditions, placed symmetrically around $\frac{\pi}{4}$.
$\epsilon$ is the off-diagonal coupling, and $N_s$ is the dimension of the Hamiltonian.
Note that, due to the lack of a naturally given spatial scale (distance between lattice sites),
we choose the spatial sample frequency or sampling rate of the Fourier transformation $f_s = 1$.
Together with the given size, i.e., total number of sites $N_s$, of our lattice, this defines all resulting parameters, such
as the frequency binning $\Delta f = f_s/N_s$, which corresponds to the smallest frequency, and the 
step size between neighboring lattice points $1/f_s$. Due to symmetry, only positive frequencies are shown.}
\label{Fig:3}
\end{figure}

\begin{figure}[ptb!]
\centering
\hspace*{-1cm} \includegraphics[width=10cm,height=5cm]{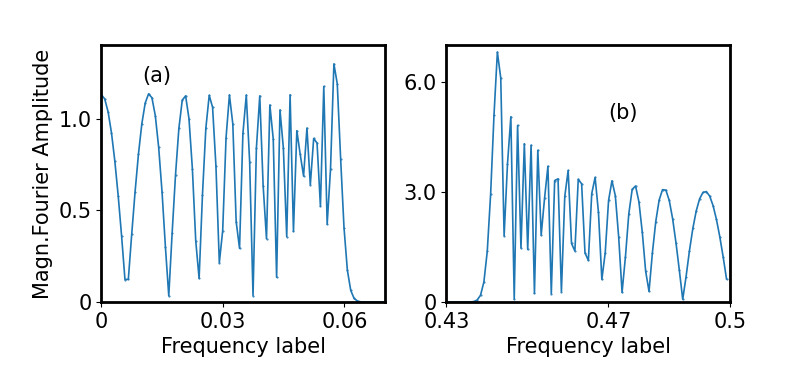}
\caption{Windows of the frequency distribution (Fourier transformation) 
for the low and high frequency domains for the individual $105$-th eigenstate (this is a state close to the transition from
localization to delocalization behavior) for an equidistant $\phi$ lattice for $d_1=1,d_2=2,\epsilon=0.01,
N_s=1202$ with open boundary conditions, placed symmetrically around $\frac{\pi}{4}$.
$\epsilon$ is the off-diagonal coupling, and $N_s$ is the dimension of the Hamiltonian.}
\label{Fig:4}
\end{figure}

\noindent
Fig.\ref{Fig:3} shows the amplitude distributions in Fourier space for the $6,70,130,213,302,416,480,550,600$-th eigenstates
corresponding to subfigures (a-i): they are representative examples covering the first band.
For the energetically low-lying localized eigenstates, and in particular for the eigenstate $6$ (see Subfig.
\ref{Fig:3}(a)), we observe two well-separated frequency distributions of finite width being
located at the edges of the finite frequency window, i.e., neighboring to the minimal and maximal possible frequencies.
Zooming into the frequency distribution close to $0.5$ one observes a sequence of peaks with a small non-constant spacing 
(not visible on the scale of Fig.\ref{Fig:3}(a))
being more than an order of magnitude smaller than the main peaks frequency close to the value $0.5$.
This behavior is illustrated in Fig.\ref{Fig:4} (a,b) for the $105$-th state for presentation
purposes, where, for both frequency
distributions at the edges of the frequency window, a progression of individual peaks with decreasing
frequency spacing is observed, reminiscent of a chirp. 

\noindent
The high frequency distribution is responsible for the short distance oscillations as well as
the much longer distance envelope behavior of the eigenstates. The latter
emerges due to the combination of the high frequency components with a small spacing
(see above discussion), similar to the well-known beating effect.
Additionally, the envelope behavior varies with increasing degree of eigenstates
when transitioning from the energetically low-lying
localized states to the more extended localized eigenstates: for extended localized eigenstates 
it acquires additional modulations. The low frequency distribution in the neighborhood of zero frequency
is responsible for this modification of the envelope structure. The spacings of the series of
peaks of this low frequency distribution matches approximately the spacing and variety
of spacings in the high frequency distribution (see Fig.\ref{Fig:4}(a,b)).
The overall envelope behavior therefore traces back to both the high and low frequency components.

\noindent
When increasing the degree of excitation, see specifically the $70-$th eigenstate in Fig.\ref{Fig:3}(b),
each of the well-separated frequency distributions in the neighborhood of the boundaries of the
overall frequency interval shows an increasing width with many more individual peaks on top of the
corresponding background. For the state $130$, we observe the onset of the separation of these frequency
distributions from the boundaries of the global frequency interval. Now each of the two distributions
possesses a substantial width. With further increasing degree of excitation, these distributions 'move'
towards the center of the frequency interval while becoming narrower, as can be seen in Fig.\ref{Fig:3}(d) 
for the $213$-th state. Finally, for the $302$-th state, which is the eigenstate at the center of
the (first) band, two extremely narrow single peaks are encountered, thereby indicating that only two frequencies
are forming the eigenstate. A closer inspection reveals that the larger (higher frequency) peak is indeed
essentially only a single frequency, whereas the smaller (lower frequency) peak consists of a few relevant
frequencies. Further increasing the degree of excitation away from the band center, the frequency distributions
broaden again and develop, step by step, an increasing number of peaks on top of a background, see the
$416$-th eigenstate in Fig.\ref{Fig:3}(f). These still separated distributions in frequency space move towards the
center of the shown frequency interval, where they start to overlap.
For the $480,550$-th eigenstates, localization sets in, and we observe a single frequency distribution which
narrows with increasing degree of excitation. Finally, for the $600$-th state, this one distribution
has become a single narrow peak, which demonstrates that only a very few neighboring frequencies contribute 
that form this peak. We remark that, for the series of localized states approaching the upper band edge,
we have only a single frequency distribution centered around the midpoint of the frequency interval,
and consequently we observe no extra modulations of the envelope structure.
Note that the sharp peak for the $600-$th eigenstate is at half the value of the maximal frequency encountered
for the state 0. 

\begin{figure}[ptb!]
%\begin{minipage}[t]{1.0\textwidth}
\hspace*{-0.6cm} % Adjust this value to shift the plot left or right
\includegraphics[width=9.5cm,height=4cm]{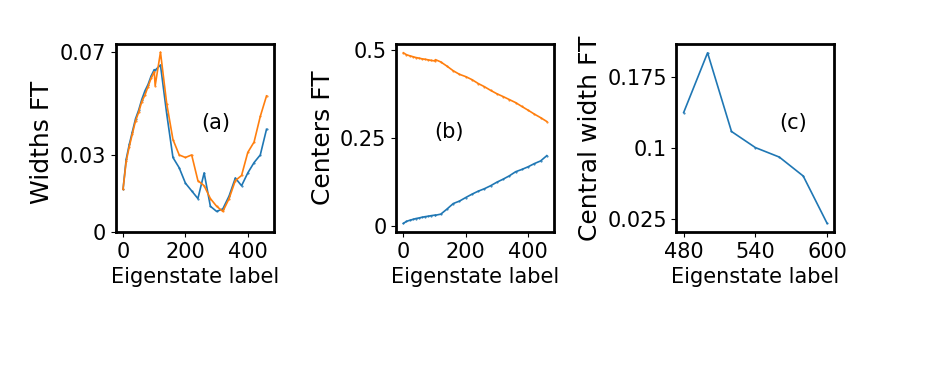}
%\end{minipage}
%\captionsetup{skip=-20pt} 
\caption{Analysis of the Fourier spectrum of the complete set of eigenstates of the first band of the
IPL for $d_1=1,d_2=2,\epsilon=0.01, N_s=1202$ with open boundary conditions,
placed symmetrically around $\frac{\pi}{4}$.
$\epsilon$ is the off-diagonal coupling, and $N_s$ is the dimension of the Hamiltonian.
(a) the evolution of the widths of the two separate frequency distributions with increasing
degree of excitation of the eigenstate, in the regime where they are still separated.
(b) The same but for the centers of the two distributions.
(c) The evolution of the width of the central single distribution, once the two distributions have merged,
i.e., for higher eigenstate label.}
\label{Fig:5}
\end{figure}

\noindent
As a next step, let us quantify the results of the Fourier analysis. To this end, we introduce the widths
and centers of the individual distributions of the Fourier transform of the eigenstates
as a function of their degree of excitation. This is shown in Fig.\ref{Fig:5}, where the
subfigure Fig.\ref{Fig:5}(a) shows the widths of the two separated frequency distributions 
as a function of the degree of excitation of the eigenstates ranging from the ground state
to the case where the two subdistributions merge ($\approx 480$-th eigenstate). 
As indicated above, starting from the ground state, we observe for both distributions an increasing
width up to a maximum, which is approximately occurring at the crossover from localized to 
delocalized states.

\noindent
Thereafter, both widths strongly decrease, pass a minimum, before they
increase again. This reflects the widening and narrowing of the two separated frequency distributions
observed at hand of specific examples in Fig.\ref{Fig:3}. It is important to note that a closer
look reveals that this transition from two to a single distribution is truly non-smooth in the
sense that many individual peaks emerge and disappear, tails and wings of the distributions appear 
and reappear in the course of this structural rearrangement process of the eigenstates,
which reflects itself on the coarse-grained level of Fig.\ref{Fig:5} (a,b,c) as a non-smooth behavior.
This holds in particular with increasing degree of excitation in the domain of delocalized states where
a substantial background develops and long-range tails of small amplitude are encountered. 

\noindent
Fig.\ref{Fig:5}(b) shows the evolution of the centers of the two separated frequency distributions: it is a simple
monotonic increase of approximately linear character. In Fig.\ref{Fig:5}(c),
the width of the one remaining centered frequency distribution for excitations higher than $480$ are
shown: apart from an initial increase there is an overall decrease by approximately an order of magnitude
resulting in a narrow peak for the eigenstate at the upper band edge of the (first) band. 
All of this shows how the 
frequency content of the eigenstates changes with increasing degree of excitation: Generally speaking,
low frequency distributions slowly move to higher frequency distributions, and the width of the distributions
oscillate between large and small values, i.e., we encounter in an oscillatory manner
broad distributions of many peaks and narrow peaks.
The main difference between the series of localized states at the
lower band edge and the upper band edge is that for the lower band edge series we have two frequency
distributions which are well-separated, whereas there is only a single higher frequency distribution 
centered at half of the maximum frequency for the series of localized states at the upper band edge.

\section{Spin Textures of the Eigenstates}
\label{spin}

\noindent
Having analyzed the quasi-nodal structure and the spectral evolution of the eigenstates across the band,
we now return to the spin interpretation of our IPL Hamiltonian. It is then natural to explore the spin
textures of the eigenstates with increasing degree of excitation
ranging, exemplarily, from the ground state to higher energy eigenstates at the center of the band. 
We will see that the spin excitation structure is closely related to the previously discussed quasi-nodal
structure in the low-energy regime of localized states, i.e., there exist strong spin-spatial correlations.
Before we enter into this analysis, let us first introduce the spin setting of our model.

\subsection{Basic Concepts and the Decoupled Case $\epsilon = 0$}
\label{spindec}

\noindent
We have shown that the IPL possesses the interpretation of a spin in an inhomogeneous rotating magnetic field
combined with a spin-dependent hopping process. To understand the spin structure of the eigenstates, let us first consider
the simple zero coupling case $\epsilon = 0$, where our IPL turns into a set of $N$ decoupled two by two
subsystems, each characterized by an angle $\phi_m$. The eigenvalues (after subtracting the offset $\frac{d_1+d_2}{2}$)
are, for each subsystem, $\pm \frac{d_1-d_2}{2}$,
and the corresponding eigenvectors are $(\text{cos} \phi_m, - \text{sin} \phi_m )$
and $(\text{sin} \phi_m, \text{cos} \phi_m )$, respectively. That means, particularly for the values
of the angle $\phi_m = 0,\frac{\pi}{4},\frac{\pi}{2}$, we have the pairs of eigenvectors
$((1,0);(0,1))$, $( \frac{1}{\sqrt{2}} (1,-1);\frac{1}{\sqrt{2}}(1,1))$ and $((0,-1);(1,0))$, respectively.

\begin{table}[H]
  \begin{center}
    \begin{tabular}{|c|c|c|c||c|c|c|} 
      \hline
      $\phi_m$ & $\langle\sigma_x \rangle$ & $\langle \sigma_y \rangle$ & $\langle \sigma_z \rangle$
      & $\langle\sigma_x \rangle$ & $\langle \sigma_y \rangle$ & $\langle \sigma_z \rangle$  \\
      \hline
      0 & 0 & 0 & 1 & 0 & 0 & -1 \\
      \hline
      $\frac{\pi}{8}$ & $-\frac{1}{\sqrt{2}}$ & 0 & $\frac{1}{\sqrt{2}}$
      & $\frac{1}{\sqrt{2}}$ & 0 & $-\frac{1}{\sqrt{2}}$\\
      \hline
      $\frac{\pi}{4}$ & -1 & 0 & 0 
      & 1 & 0 & 0\\
      \hline
      $\frac{3\pi}{8}$ & $-\frac{1}{\sqrt{2}}$ & 0 & $-\frac{1}{\sqrt{2}}$
      & $\frac{1}{\sqrt{2}}$ & 0 & $\frac{1}{\sqrt{2}}$\\
      \hline
      $\frac{\pi}{2}$ & 0 & 0 & -1
      & 0 & 0 & 1\\
      \hline
    \end{tabular}
  \end{center}
    \caption{Expectation values of the different spin operators for certain blocks on the diagonal
    of the Hamiltonian for $\epsilon = 0$, taken for the eigenstates with the eigenvalues $\frac{d_1-d_2}{2}$,
    columns two to four, and $\frac{d_2-d_1}{2}$, columns five to seven. The corresponding eigenvectors are
    ($\text{cos} \phi_m,-\text{sin} \phi_m$) and ($\text{sin} \phi_m,\text{cos} \phi_m$),
     specified by the angle $\phi_m$.}
   \label{table:1}
\end{table}

\noindent
For each two by two subsystem we have a local set of spin operators $\sigma_x^m,\sigma_y^m,\sigma_z^m$,
which allows us to define the local spin expectation values 
$\langle \psi^m | \sigma_i^m | \psi^m \rangle, m = 1,...,N; i \in x,y,z$, where $|\psi^m \rangle$ is the
local eigenvector on subsystem/cell $m$. This provides us with the local spin polarization.
We note that the expectation values of $\sigma_y^m$ are always zero 
due to our real symmetric Hamiltonian. For the eigenvalue $\frac{d_1-d_2}{2}$
with the eigenvector ($\text{cos} \phi_m,-\text{sin} \phi_m$) and taking the values $\phi_m = 0, \frac{\pi}{8},
\frac{\pi}{4},\frac{3\pi}{8},\frac{\pi}{2}$, the $\sigma_x$-polarization ranges from $0$ via $-1$ to $0$ again,
and the $\sigma_z$-polarization ranges from $1$ via $0$ to $-1$, see the left half of table \ref{table:1}.
For the eigenvalue $\frac{d_2-d_1}{2}$
with the eigenvector ($\text{sin} \phi_m,\text{cos} \phi_m$) and taking the same values for $\phi_m$,
the $\sigma_x$-polarization ranges from 0 via 1 to 0,
and the $\sigma_z$-polarization goes from -1 via 0 to 1, see the right half of table \ref{table:1}.
Overall, therefore, the spin expectation
values of one of the eigenstates can simply be obtained from the other one by multiplying by $(-1)$, i.e.,
the spin expectation values are inverted.

\subsection{Spin Structure of the IPL}
\label{spinipl}

\begin{figure*}[t]
\includegraphics[width=19cm,height=11cm]{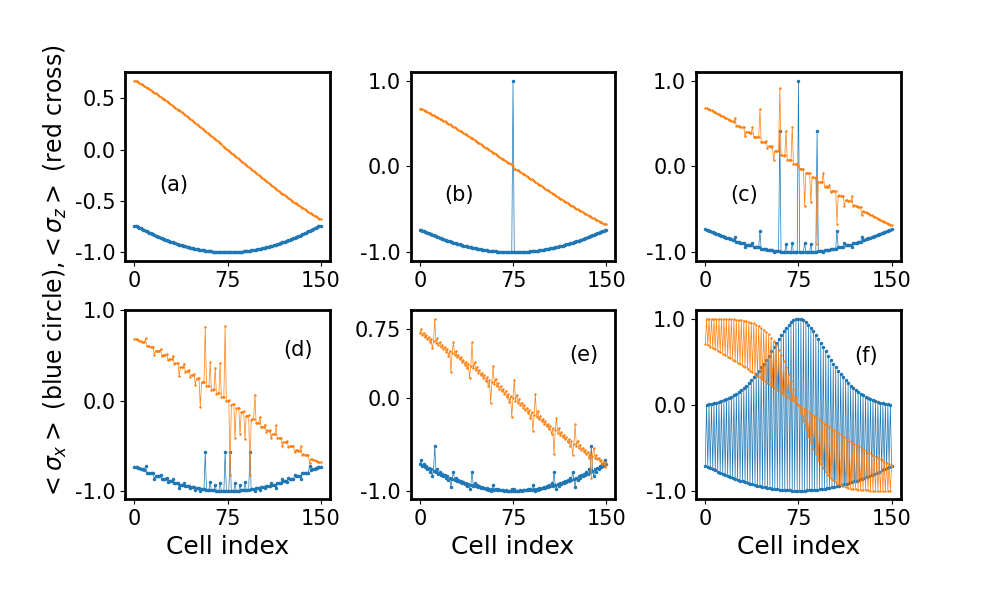}
\caption{Local spin structure, i.e., local expectation values of $\sigma_x$ and $\sigma_z$ of individual
eigenstates of the IPL for $d_1=1,d_2=2,\epsilon=0.01,
N_s=302$ with open boundary conditions, placed symmetrically around $\frac{\pi}{4}$.
$\epsilon$ is the off-diagonal coupling, and $N_s$ is the dimension of the Hamiltonian.
(a-f) correspond to the $0,1,19,30,70,75$-th eigenstate ranging from the ground
state to the eigenstate in the center of the first band, respectively.
The cell index labels the cells up to the total number of $2\times2$ cells of the lattice, i.e., half 
the total number of its sites $N_s$.}
\label{Fig:6}
\end{figure*}

\noindent
Let us now explore the spin structure of the eigenstates with increasing degree of excitation.
We employ the local spin expectation values for given eigenstates, which are defined as follows

\begin{eqnarray}
\langle \sigma_x \rangle_m = \langle \Psi^m | \sigma_x | \Psi^m \rangle \hspace*{1cm} m = 1,...,N_s/2, \\ \nonumber
\langle \sigma_z \rangle_m = \langle \Psi^m | \sigma_z | \Psi^m \rangle \hspace*{1cm} m = 1,...,N_s/2, 
\end{eqnarray}

\noindent
where $|\Psi^m \rangle$ is now based on the exact eigenvector of the IPL: it represents 
the ket of a normalized vector that consists of the two components of the eigenvector 
of the IPL living in the m-th $2\times2$ cell,
where the number of cells is half the dimension $N_s$ of the lattice Hamiltonian.
Concerning the spin textures shown in Fig.\ref{Fig:6}, we use the same selection of
eigenstates whose spatial distributions are shown in Fig.\ref{Fig:1}, which allows us to
compare them directly. We remark that in the following we omit the extra index $m$ for the local
spin expectation values in cases it is obvious in the context of the corresponding discussion.

\noindent
Let us state ahead of our analysis that there is a stunning variety of spin textures
accompanying the bands of eigenstates. We begin with the ground state analysis, whose spin profile
and eigenvectors are shown in Fig.\ref{Fig:6}(a) and Fig. \ref{Fig:1}(a), respectively.
Note that the cell index labels consecutively the cells of the lattice.
The ground state is Gaussian-like \cite{Schmelcher25} and essentially localized between the lattice sites $100$ and $200$,
as shown in Fig. \ref{Fig:1}(a). 
The spin texture of the ground state varies smoothly and quadratically around the extremal value 
$\langle \sigma_x \rangle = -1$ and 
linearly, thereby crossing zero from positive to negative values, around the value $\langle \sigma_z \rangle = 0$. 
This is in accordance with what one might expect from the lowest energy eigenstate, see the 
$\epsilon = 0$ case in table \ref{table:1}.

\noindent
Let us next inspect the first excited state, as shown in Fig.\ref{Fig:1}(b).
As discussed in section \ref{quasinodes}, it shows an envelope with a quasi-nodal structure at the center of the lattice. 
Right at the position of this quasi-nodal structure, the corresponding
spin texture now shows for $\langle \sigma_x \rangle$ a spin peak involving a change of the corresponding
local expectation value from $-1$ to $+1$. At the same location $\langle \sigma_z \rangle$ shows a minor dip.
Therefore, this first excited state discriminates itself from the ground state by a very localized spin excitation,
which plausibly costs little energy because it happens close to the quasi-node of the eigenstate.
This behavior persists when increasing the degree of excitation, i.e., considering higher excited eigenstates.
Indeed, moving on to the $n-th$ excited state in the regime of localized states, we observe for the envelope
$n$ quasi-nodal structures. Correspondingly, the spin texture shows pairs of spin peaks for $\langle \sigma_i \rangle, i=x,z$
with varying amplitudes that depend on the details of the couplings within the cells.
Odd states typically possess a large central spin peak of $\langle \sigma_x \rangle$.
Fig.\ref{Fig:1}(c) and Fig.\ref{Fig:6}(c) show the eigenvector component distribution and the spin profiles for the
$19-$th eigenstate, which is a state at the beginning of the crossover from localization to delocalization.
We encounter a series of spin peaks for $\langle \sigma_i \rangle, i=x,z$
on top of the corresponding 'smooth' backgrounds given by the ground state that are located at the
positions of the $19$ quasi-nodes. The height and signs of these peaks are largely varying, but the
general tendency can be observed that the closer the quasi-nodal structure approaches the value zero,
the larger the spin peaks are, see also subsection \ref{spinspace}, with a marked difference in the behavior of
$\langle \sigma_x \rangle$ vs. $\langle \sigma_z \rangle$.

\noindent
Fig.\ref{Fig:1}(d) and Fig.\ref{Fig:6}(d) show the eigenvector component distribution and the spin profiles for the
$30-$th eigenstate, which is a delocalized state. Now the length of the segments of the background of the 
spin texture between neighboring spin peaks has decreased substantially, and the spin fluctuations are
quite irregular (see also Fig.\ref{Fig:5} in subsection \ref{fourier} signifying this transition in the Fourier spectrum).
We are here in the regime of increased excitation energies, where the low-energy envelope behavior is absent.

\noindent 
Regular behavior reemerges when approaching the center of the band, as can be seen by inspecting
Fig.\ref{Fig:1}(e) and Fig.\ref{Fig:6}(e), showing the eigenvector component distribution and the spin profiles for the
$70-$th eigenstate. The eigenstate now shows a well-structured regular and repetitive spin texture.
The larger spin peaks of both character $\sigma_x,\sigma_z$
are arranged almost equidistantly, with small amplitude peaks in between. This corresponds 
to the regime where the underlying Fourier distribution shows a strong narrowing (see Fig.\ref{Fig:3}(e)). 
This culminates in the $75-$th eigenstate at the center of the band (see Fig.\ref{Fig:1}(f) and Fig.\ref{Fig:6}(f)),
which shows a staggered, highly regular spin configuration from cell to cell across the complete lattice,
associated with an extremely narrow Fourier spectrum of the eigenvector component distribution.
In the spirit of the above observations, the IPL shows a transition from a quasi-continuous, quasi-smooth
low-energy and large-scale behavior (apart from neighboring site oscillations)
to a heavy oscillatory short-scale behavior in the center of the band.
The crossover between these two regimes involves a certain degree of irregularity
quantified above by the large width of the underlying Fourier distribution, whereas these two regimes themselves
are much more regular, as reflected by the small width of the corresponding Fourier peaks, see subsection
\ref{fourier}.

\noindent
Let us now analyze the eigenstates' global spin fluctuations quantitatively. To this end, we employ the total
variation (TVS) of the local spin expectation values $\langle \sigma_x \rangle_m, \langle \sigma_z \rangle_m$
across each eigenstate. We inspect the behavior of this TVS with varying eigenstates
across the complete spectrum. It is defined as 

\begin{eqnarray}
\langle \sigma_x \rangle_{tv} = \sum_{m=1}^{N_s/2-1} |\langle \sigma_x \rangle_{m+1} - \langle \sigma_x \rangle_{m}|, \\ \nonumber
\langle \sigma_z \rangle_{tv} = \sum_{m=1}^{N_s/2-1} |\langle \sigma_z \rangle_{m+1} - \langle \sigma_z \rangle_{m}|.
\end{eqnarray}

\begin{figure}[ptb!]
\centering
\includegraphics[width=10cm,height=6cm]{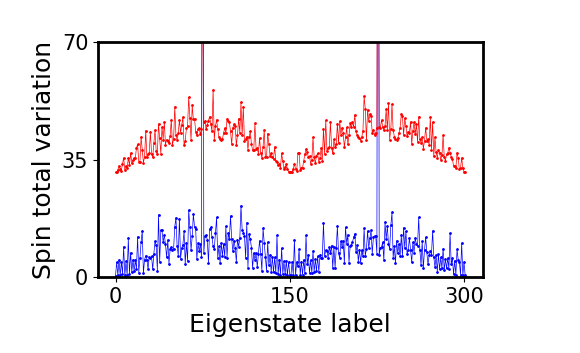}
\caption{Total variation of the local spin expectation value of each eigenstate across the eigenstate spectrum,
i.e., with increasing degree of excitation, for both bands. Red upper curve corresponds to 
$\langle \sigma_z \rangle$ and the blue lower curve
to $\langle \sigma_x \rangle$. Equidistant $\phi$ lattice for $d_1=1,d_2=2,\epsilon=0.01, N_s=302$
with open boundary conditions, placed symmetrically around $\frac{\pi}{4}$.
$\epsilon$ is the off-diagonal coupling, and $N_s$ is the dimension of the Hamiltonian. Note that the 
$\langle \sigma_z \rangle$ red curve has been shifted by a constant value of $30$ for reasons of clarity.}
\label{Fig:7}
\end{figure}

\noindent
Fig.\ref{Fig:7} shows the TVS for the complete spectrum of eigenstates across both bands.
We observe that both spin values $\langle \sigma_x \rangle_{tv}, \langle \sigma_z \rangle_{tv}$
show, at first glance, an overall very similar behavior. They exhibit two dominant narrow central peaks each.
These dominant peaks correspond to the eigenstates at the band centers (see, e.g., Fig.
\ref{Fig:1}(f)) whose spin profile shows the highest possible frequency and strong oscillatory behaviors
for both spin components (see Fig.\ref{Fig:6}(f)). This is why this TVS acquires a much larger
value compared to the TVS of the other eigenstates.

\noindent
The envelope (averaged) behavior of the TVS across both bands consists of two oscillations,
the amplitude of the $\langle \sigma_z \rangle_{tv}$ oscillation being larger than that of the 
$\langle \sigma_x \rangle_{tv}$ oscillation.
The maxima of the envelopes correspond to the centers of the bands. This oscillatory
behavior corresponds to the fact that, starting from the ground state, the number of spin peaks increases
with increasing degree of excitation of the eigenstates (see Figs.\ref{Fig:1},\ref{Fig:6}),
which makes up for an increase of the corresponding total variation. Approaching the center of the band,
the (averaged) TVS behavior flattens out until it is maximal at the center of the band, and subsequently 
decreases again with further increasing degree of excitation. These statements hold
independently of whether we have localized or delocalized states. 

\begin{figure}[ptb!]
\centering
\hspace*{-0.5cm}
\includegraphics[width=10cm,height=6cm]{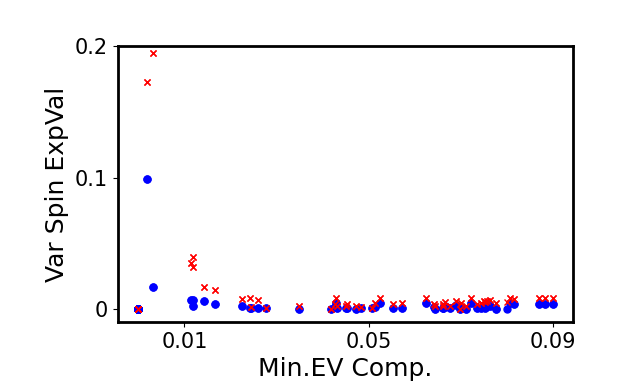}
\vspace*{-0.4cm}
\caption{Spin space scatter plot of the $12$-th eigenstate
of the IPL for $d_1=1,d_2=2,\epsilon=0.01,N_s=302$
with open boundary conditions, placed symmetrically around $\frac{\pi}{4}$.
$\epsilon$ is the off-diagonal coupling, and $N_s$ is the dimension of the Hamiltonian.
Shown are the variation of the local spin expectation values for $\sigma_x$ (blue dots) and
$\sigma_z$ (red crosses) versus the minimal eigenvector component (magnitude) 
for each cell.} 
\label{Fig:8}
\end{figure}

\noindent
Looking into the more detailed structure we observe that there is high frequency
irregular oscillations for the TVS with increasing degree of excitation.
The TVS of $\langle \sigma_x \rangle$ shows more such
high frequency oscillations (oscillations between neighboring degrees of excitation) 
as compared to the TVS of $\langle \sigma_z \rangle$, which, typically, possesses oscillations of a
somewhat smaller frequency. 

\begin{figure*}[t]
\includegraphics[width=19cm,height=11cm]{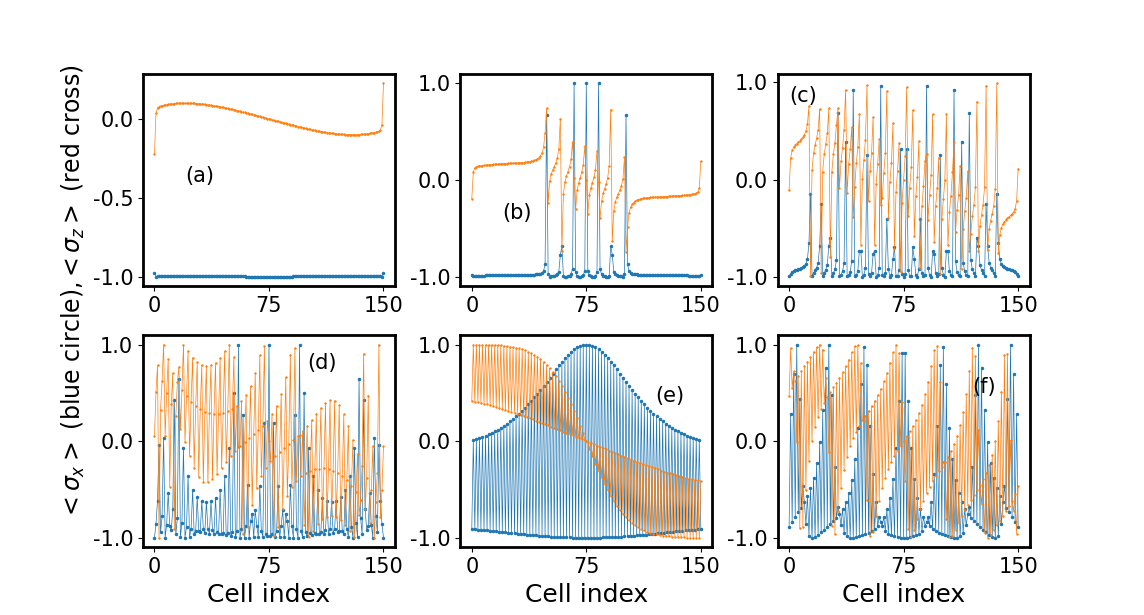}
\caption{Local spin structure, i.e., local expectation values of $\sigma_x$ and $\sigma_z$ of individual
eigenstates of the IPL for $d_1=1,d_2=2,\epsilon=0.3,
N_s=302$ with open boundary conditions, placed symmetrically around $\frac{\pi}{4}$.
$\epsilon$ is the off-diagonal coupling, and $N_s$ is the dimension of the Hamiltonian.
(a-f) correspond to the $0,7,26,49,75,78$-th eigenstate, respectively.
The cell index labels the cells up to the total number of $2\times2$ cells of the lattice, i.e., half 
the total number of its sites $N_s$.}
\label{Fig:9}
\end{figure*}

\subsection{Spin-Spatial Correlations}
\label{spinspace}

\noindent
In subsection \ref{spinipl}, we have observed that the location of the quasi-nodes of the eigenstates correspond
to that of the spin peaks. Let us now work out the spin spatial correlations in some more detail.
To this end, we use a scatter plot for the eigenstates, which shows the variation of the
local spin expectation values $\langle \sigma_i \rangle_m, i = x,z$ versus the minimal eigenvector component (magnitude)
for each cell. Fig.\ref{Fig:8} shows the scatter plots for these two quantities for the $12-$th eigenstate
of the first band, which corresponds to a localized state. It shows that large values
of the spin variation corresponding to large spin peaks occur only for small values of the eigenvector components,
establishing a unique correlation behavior. This is in correspondence with the
above observation that the quasi-nodes position corresponds to the positions of the spin peaks.
With increasing cell eigenvector components, the corresponding spin variations and peak heights decrease
rapidly and fluctuate in a small interval close to zero.
These observations are of generic character for the localized states but extend somewhat into the regime of
delocalized states.

\section{The Stronger Coupling Case}
\label{coupling}

\noindent
In the previous sections we have been focusing on the weak coupling case $\epsilon = 0.01$.
Many of the observations made for the latter case hold actually for a broad window of coupling
strengths.
This is why we focus in this section only on the major differences
for much stronger couplings at hand of the case $\epsilon = 0.3$.

\noindent
Fig.\ref{Fig:9} shows the spin textures $\langle \sigma_i \rangle, i=x,z$ for the 
$0,7,26,49,75,78$-th eigenstates of a lattice with $N_s=302$, ranging from the ground
state to states close to the center of the first band. Similar to the case $\epsilon = 0.01$, we encounter
regimes of localized states that show a corresponding envelope and possess quasi-nodes,
and the local expectation values of the spins peak at those quasi-nodes.
However, there is some marked difference for the $\epsilon = 0.3$ case compared to the $\epsilon = 0.01$ case.
For the ground state, see Fig.\ref{Fig:9}(a), we encounter an expectation value $\langle \sigma_x \rangle$
that is approximately constant across the complete lattice, whereas $\langle \sigma_z \rangle$
varies smoothly from positive to negative values. Fig.\ref{Fig:9}(b,c) show the spin texture
for the $7,26$-th excited eigenstate: the spin peaks of $\langle \sigma_x \rangle$ are now almost
equidistantly arranged according to the quasi-nodal structure of the corresponding eigenstates.

\begin{figure}[ptb!]
\centering
\includegraphics[width=10cm,height=6cm]{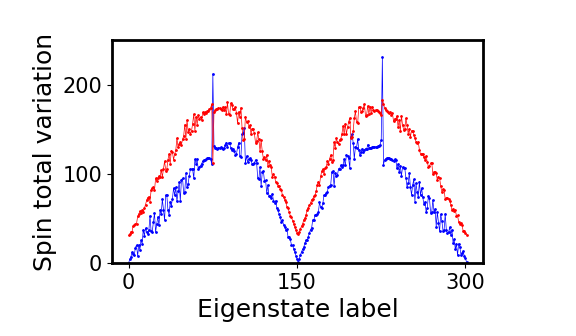}
\caption{Total variation of the local spin expectation value of each eigenstate across the eigenstate spectrum,
i.e., with increasing degree of excitation, for both bands. Red upper curve corresponds $\sigma_z$ and the blue lower curve
to $\sigma_x$. IPL for $d_1=1,d_2=2,\epsilon=0.3, N_s=302$
with open boundary conditions, placed symmetrically around $\frac{\pi}{4}$.
$\epsilon$ is the off-diagonal coupling, and $N_s$ is the dimension of the Hamiltonian.
Note that the 
$\langle \sigma_z \rangle$ red curve has been shifted by a constant value of $30$ for reasons of clarity.}
\label{Fig:10}
\end{figure}

\noindent
They occur on top of an almost flat background and exhibit predominantly a large amplitude.
The above holds in the regime of localized states. Moving on to the regime of delocalized
states, the spin patterns strongly rearrange, as can be seen in Fig.\ref{Fig:9}(d) for the $49$-th excited
eigenstate, and they exhibit a combination of (ir)regular modulations. 
Close to, and in particular right at, the center of the (first) band, regularity takes over
again, see the staggered spin configuration in Fig.\ref{Fig:9}(e) for the $75$-th eigenstate
and the nearby $78$-th eigenstate spin texture in Fig.\ref{Fig:9}(f). 

\noindent
Fig.\ref{Fig:10} shows the corresponding TVS for $\epsilon = 0.3$.
As indicated above, the spin peaks are much more pronounced, which leads
to the fact that now the individual humps of the total variation are much larger in amplitude, and the main
central peaks due to the eigenstates at the band centers are much less dominant. 
Also, the relative fluctuations on top of the envelope behavior are much smaller as compared to the $\epsilon = 0.01$ case.
Generally speaking, the fluctuations in the first half of the first band and the second half of the
second band are significantly larger compared those of the 'inner' parts of the bands.

\section{Conclusions and Outlook}
\label{concl}

\noindent
Isospectrally patterned lattices possess some remarkable properties, such as the tunable composition
of their energy bands in terms of localized versus delocalized eigenstates. While the first works
along these lines \cite{Schmelcher25,Schmelcher26} lack an immediate physical interpretation of
the IPL Hamiltonian, we provide here a very natural setting of the case of a real symmetric
tight-binding situation composed of coupled isospectral two-site cells. Our inhomogeneous finite lattice
Hamiltonian can be interpreted in terms of a single spin that is exposed to a rotating magnetic field and is
allowed to hop in a spin-flipping manner from cell to cell. Since no static magnetic field configuration
does provide this 'globally', an experimental realization would rely on a slice through an inhomogeneous magnetic
field configuration. For example, at a constant distance from a current-carrying wire, the field rotates when moving
across a (part of a) circle. The general setting would then be ultracold spin one half fermions
in an optical lattice which could tunnel from site to site and are superimposed with the above-mentioned
rotating magnetic field. Spin-flip hopping could be achieved by Raman-assisted or spin-orbit coupled
tunneling. This is well within reach in terms of current ultracold
quantum gas experiments \cite{Bloch08,Tarruell18,Bakr25,Esslinger10,Lewenstein17}.
Even more, single site resolution quantum gas microscopes in optical lattices \cite{Bakr09,Sherson10,Buob24}
allow to monitor, in particular, eigenstate densities in many-shot experiments. Alternatively to the above,
and without relying on an external magnetic field, one could use double
well superlattices, where a single double well constitutes a pseudospin one-half, and the IPL engineering
of the diagonal blocks of the lattice would correspond to a site by site adjustment of the well-depths
and of the off-diagonal couplings \cite{Rey07,Gao22}. 

\noindent
Inspired by its spin interpretation, we have gone in the present work one step ahead in the analysis of the 
IPL. The localized eigenstates in the vicinity of the low and high energy band edges show an oscillatory envelope structure 
with quasi-nodes separating the individual oscillations. The distribution of these quasi-nodes
follows a characteristic pattern with an increasing distance between the quasi-nodes with increasing distance
from the lattice center. The corresponding Fourier spectrum of the low-energy eigenstates is a bimodal distribution
with well-separated subdistributions at the lower and upper frequency edge, whose widths increase with
increasing degree of excitation. These subdistributions consist of a sequence of well-separated peaks
that accumulate towards the edge of the distribution. Remarkably, it turns out that excited eigenstates exhibit spin peaks
at the positions of their quasi-nodes, which are unique features of the degree of
excitation and can be identified by inspecting the spin-spatial correlations.
This way, characteristic spin patterns emerge while moving across the spectrum of energy eigenstates
in the band. The total spin variation of an eigenstate has proven to be a helpful quantity to showcase
the variability within this spectrum. 

\noindent
While passing from the regime of localized to the regime of delocalized eigenstates, we enter a crossover
regime where the envelope behavior, as far as possible, dissolves, and both the eigenstate profiles as well
as the spin textures rearrange. The Fourier frequency distributions leave the edges of the available frequency
window and move towards the center position of the frequency window 
while narrowing their widths. Finally, when approaching the center
of the band, two extremely narrow peaks form. At the same time, the spin texture becomes highly regular until
a highly oscillatory staggered spin configuration emerges for the band center eigenstate. This beautifully demonstrates
the rich spin textures realized by the IPL. We have quantified the evolution of the Fourier spectrum by tracing
the widths and centers of the Fourier distributions with varying degree of excitation in the band.
Considering the eigenstates in the spectrum from the center of the band to the upper band edge,
again a bimodal distribution emerges with significant widths. The two subdistributions now move
both towards the center of the overall frequency window, start to overlap, and finally form a single
distribution that subsequently narrows to a single sharp peak for the eigenstate at the upper band edge.

\noindent
While the above demonstrates the variability and richness of the spectral behavior of the IPL considered
here, one has to keep in mind that the present study deals only with a special, i.e., the most simple, case
of the broader class of IPL setups. Indeed, generalizations could happen along two main lines: (i) increase
the dimensionality of the degenerate subspace from two to higher dimensions along with the dimensionality
of the orthogonal or unitary transformations and their angles/phases, (ii) explore the case of higher dimensional
angle/phase grids going into two or three-dimensional lattices. From what we know about the present case,
expectations are significant that there will be many more intriguing properties and phenomena to be observed.

\section{Acknowledgments}
\label{ACK}

The author acknowledges helpful discussions with F.K. Diakonos and D. Walle. Technical support by A. Becker
is appreciated very much.

\end{document}